# Effect of interdiffusion and quantum confinement on Raman spectra of the Ge/Si(100) heterostructures with quantum dots

I.V Kucherenko[1], V.S. Vinogradov[1], N.N. Melnik[1], L.V. Arapkina[2], V.A. Chapnin[2], K.V. Chizh[2], V.A. Yur'ev[2]
[1]Lebedev Physical Institute RAS, 119991 Moscow, Leninsky pr. 53, Russia
[2]Prokhorov General Physics Institute RAS, 119991 Moscow Vavilova str. 38, Russia
E-mail: kucheren@sci.lebedev.ru; vvs@sci.lebedev.ru

**Abstract.** We used Raman scattering for study the phonon modes of self-organized Ge/Si quantum dots, grown by a molecular-beam-epitaxy method. It is revealed, that Ge-Ge and Si-Ge vibrational modes considerably intensify at excitation of exciton between the $\Lambda_3$ valence and $\Lambda_1$ conduction bands (transitions $E_1$ and $E_1 + \Delta_1$), that allows to observe Raman scattering spectrum from extremely small volumes of Ge, even from one layer of quantum dots with the layer thickness of ~10 Å. It is shown that Si diffuses into the Ge quantum dots from the Si spacer layers forming $Ge_xSi_{1-x}$ solid solution, and Si concentration was estimated. It is revealed, that the frequency of Ge-Ge mode decreases in 10 cm$^{-1}$ at decreasing of the Ge layer thickness from 10 up to 6 Å as a result of phonon size confinement effect.

We have utilized Raman scattering (RS) to study the phonon modes in Ge/Si structures with self-organized Ge quantum dots (QDs). These structures consisting of 5 periods were grown by molecular-beam epitaxy (MBE) on a (001) p-type Si substrate with a residual gas pressure of 5×10$^{-10}$ Torr. The growth chamber was integrated with a scanning tunneling microscope (STM) GPI-300 for analysis of QD shape and size. The Ge layers with an effective thickness in the range of 4–18 Å were grown at 350°C. Each Ge layer was covered with a 50 nm undoped Si layer deposited at 530°C. According to STM investigations hut clusters are forming on Si(100) surface for the Ge layer thicknesses of 6–12 Å.

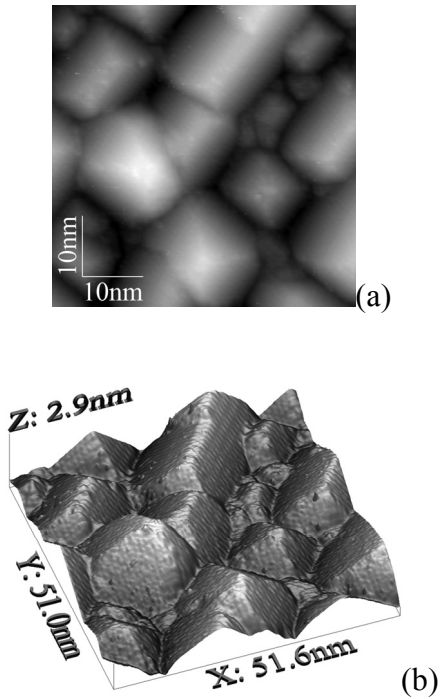

Fig. 1. Two-dimensional (a) and three-dimensional (b) STM images of Ge hut clusters on Si (001) surface; effective thickness of the Ge layers is 10 Å.

Varying the thickness of the Ge layer we can obtain structures with quantum dots of different size with height ($h$) of 0.6–1.5 nm and dimensions in the growth plane ($l$) of 6–15 nm. The QD density is ~ 5×10$^{11}$ cm$^{-2}$. Typical STM image of the sample with $h_{Ge}$ = 10 Å is presented in Fig. 1. STM image is obtained for the tunneling current $I_t$ = 0.1 nA and voltage $V_t$ = +2.1 V.
Raman spectra were excited by an Ar-ion laser with the wavelength $\lambda$ = 514.5; 488 nm and He-Cd laser with $\lambda$ = 441.6 nm. The scattered spectra were analyzed at room temperature with U-1000 spectrometer with resolution of 1 cm$^{-1}$.
It is well known that in Ge/Si nanostructures with Ge QDs three main peaks are dominated: peak at $f$ = 520 cm$^{-1}$, peak near $f$ = 300 cm$^{-1}$, and the band in the vicinity of 400 cm$^{-1}$. These frequencies are related to Si-Si, Ge-Ge and Ge-Si vibrations. Frequency dependences of Ge-Ge and Ge-Si modes and their line widths on the effective thickness of Ge layers ($h_{Ge}$) are studied in this paper. Frequencies of Ge-Ge modes in the samples with $h_{Ge} \geq$ 10 Å exceed frequency of the same mode of bulk Ge crystals by ~12 cm$^{-1}$ due to elastic strains in the growth plane. But we observed decrease of Ge-Ge mode frequencies from 312 to 301 cm$^{-1}$ while $h_{Ge}$ decreases from 10 to 6 Å (Fig. 2). The smallest line width (8 cm$^{-1}$) corresponds to $h_{Ge}$ = 9 and 10 Å. In these samples hut cluster size dispersion is minimal.
We studied influence of wetting and spacer layers on the RS spectra of the structures. For this purpose Ge/Si structure with 4 Å Ge layers but before formation of QDs (similar to wetting layers) were grown. It is shown that contribution from Ge layers in this sample into Raman spectrum can be ignored. Comparison of Raman spectra of two samples consisting of one 10 Å Ge dot layer, uncapped and caped with 50 nm Si layer (spacer), clearly shows that interdiffusion between Si spacer layer and Ge dots takes place during growth.

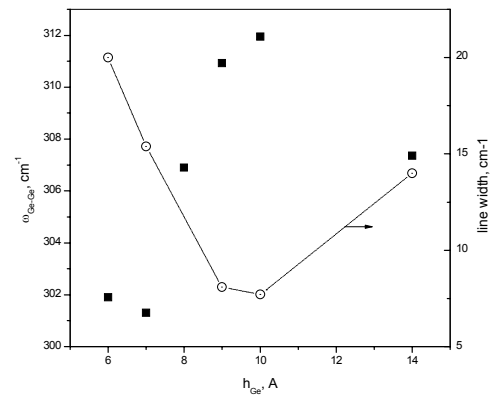

Fig. 2. Dependences of frequency and line width of Ge-Ge mode on the effective thickness of Ge layer. Full squares designate frequencies of Ge-Ge mode.

Both Ge-Ge and Ge-Si phonon modes are found to exhibit strong enhancement in capped sample, and the width of their lines is narrow (7–8 cm$^{-1}$). We relate this enhancement to strong interaction between these modes and $E_1$ exciton. Energy of $E_1$ transition ranges up to about 2.6 eV in Ge$_x$Si$_{1-x}$ solid solution at $x = 0.65$ [1]. This energy is in resonance with the laser excitation energy $E = 2.54$ eV ($\lambda = 488$ nm). According to our estimations the amount of Si in Ge QDs is about 35% due to strong Si-Ge intermixing. The dependence of the Ge-Ge mode intensity on the laser wavelength also proves a resonant character of RS, the highest intensity of this mode corresponds to $\lambda = 488$ nm. Due to resonance Raman scattering it is possible to obtain spectra even from one dot layer. It is shown that Raman spectra have resonance character in the Ge/Si structures with $h_{Ge} = 9$ and 10 Å at Si concentration inside dots about 34%. Concentration of Si inside the dots was obtained from the integral intensity ratio of Ge-Ge and Ge-Si modes [2]. Si concentration remains ~ 34% in the range of $h_{Ge} = 6$–10 Å. We observed decrease of Ge-Ge mode frequency up to 5 cm$^{-1}$ with decrease of the laser wave length (514.5, 488 and 441.6 nm). This result correlates with [3]. We explain this phenomenon by size dispersion of QDs in our samples. At high excitation energy ($\lambda = 441.6$ nm) small QDs give resonant contribution into RS spectra, and according to Fig. 2 small QDs have less Ge-Ge mode frequency.

Non-monotonic dependence of Si-Si mode intensity on the effective Ge layer thickness is observed at excitation by Ar$^+$-laser with $\lambda = 488$ nm: line intensity is the smallest in the samples in which resonance Raman scattering of Ge-Ge and Ge-Si modes occurs and absorption in Ge layers is high ($h_{Ge} = 8$, 9 and 10 Å).

Therefore the effect of different factors on RS spectra in Ge/Si structures with QDs of hut shape is analyzed: interdiffusion, elastic strain and phonon quantum confinement. It is shown that Si diffuses into Ge quantum dot from the lying above Si spacer through faces and edges of pyramid where elastic strains and concentration gradient are highest. As the concentration of Si in QDs remains approximately constant in the samples with $h_{Ge} = 6$–10 Å, and elastic strains do not depend on the size of QD [4], we came to the conclusion that reduction of the Ge-Ge mode frequency (by about 10 cm$^{-1}$) with decrease of $h_{Ge}$ in the range of 10–6 Å is due to the phonon confinement effect. Using approximation of LO [100] dispersion relation of Ge [5] in the range of wave vectors $q/q_{max} = 0$–1/2 by the expression $\omega = \omega_0 [1 - \alpha (q/q_{max})^2]$ where $q = \pi/d$, $q_{max} = \pi/a_{Ge}$, $a_{Ge}$ is Ge lattice constant, $\alpha = 8/15$, we obtained for our data $\omega_0 = 312$ cm$^{-1}$ and $\Delta\omega = -10$ cm$^{-1}$ the size of the confinement region $d = 23$ Å. This value is close to the length of cube which volume equal to pyramidal volume with $h = 6$ Å and $l = 100$ Å.

## *Acknowledgements*

The work was supported by the Basic Research Foundation (grant N 07-02-00899-a), by the Basic Research Program "Quantum Nanostructures" of RAS (No 5.4 Vin) and by the Federal Agency on Science and Innovations of the Ministry of Education and Science of RF (state contract No 02.513.11.3130).